\documentclass[12pt,showkeys]{revtex4}
\usepackage{graphics}

\begin{document}

\title{Recurrent epidemics in small world networks}
\author{J. A. Verdasca$^a$,  M. M. Telo da Gama$^{a*}$, A. Nunes$^a$, N. 
R. Bernardino$^a$, J. M. Pacheco$^a$ and M. C. Gomes$^b$ }
\address{$^a$Centro de F{\'\i}sica Te{\'o}rica e Computacional and
Departamento de F{\'\i}sica, Faculdade de Ci{\^e}ncias da Universidade de 
Lisboa, P-1649-003 Lisboa Codex, Portugal\\
$^b$Departamento de Biologia Vegetal, 
Faculdade de Ci{\^e}ncias da Universidade de Lisboa,\\Campo Grande, 
1749-016 Lisboa, Portugal
\\$^*$Corresponding author: margarid@cii.fc.ul.pt}

\begin{abstract}

The effect of spatial correlations on the spread of infectious diseases was 
investigated using a stochastic SIR (Susceptible-Infective-Recovered) 
model on complex networks. 
It was found that in addition to the reduction of the effective transmission 
rate, through the screening of infectives, spatial correlations may have 
another major effect through the enhancement of stochastic fluctuations.
As a consequence large populations will have 
to become even larger to 'average out' significant differences from the 
solution of deterministic models. 
Furthermore, time series of the (unforced) model provide patterns of 
recurrent 
epidemics with slightly irregular periods and realistic amplitudes, 
suggesting that stochastic models together with complex networks of 
contacts may be sufficient to describe the long term dynamics of some 
diseases.
The spatial effects were analysed quantitatively by modelling 
measles and pertussis, using a SEIR (Susceptible-Exposed-Infective-Recovered) 
model. Both the period and the spatial coherence of the epidemic peaks 
of pertussis are well described by the unforced model for realistic values 
of the parameters.

\end{abstract}


\maketitle

\vskip2pc

\newcommand\lsim{\mathrel{\rlap{\lower4pt\hbox{\hskip1pt$\sim$}}
    \raise1pt\hbox{$<$}}}
\newcommand\gsim{\mathrel{\rlap{\lower4pt\hbox{\hskip1pt$\sim$}}
    \raise1pt\hbox{$>$}}}

\section{Introduction}

Deterministic models of epidemic spread have been used for decades and 
developed into a mature branch of applied mathematics. Indeed these models 
proved successful in understanding some of the mechanisms determining the 
time evolution and the spread of a variety of infectious diseases, spawning 
the development of strategies of epidemic control (Anderson \& May 1991, 
Hethcote 2000, Murray 2003). 
When applied to immune-for-life diseases such as measles, these models 
account for the observed undamped oscilations registered in the 
pre-vaccination era through the introduction of seasonal forcing terms 
(Aron \& Schwartz 1984, Bolker 1993, Dietz 1976, Keeling \& Grenfell 1997, 
2002). 
Indeed, the seasonally forced SIR (Susceptible-Infective-Recovered) 
model describes the pre-vaccination data for measles semi-quantitatively, 
due to the prominence of epidemic annual/biennial cycles, corresponding to 
moderate annual forcing (Earn {\it et al.} 2000). 

Nevertheless, the importance of demographic stochasticity in the dynamics 
and 
persistence of childhood diseases has long been acknowledged, not least 
because of stochastic extinction. In work published in the 50's Bartlett 
(Bailey 1975, Bartlett 1957) has shown how a stochastic 
version of the deterministic SIR model exhibits small amplitude fluctuations, 
modulated by the underlying deterministic period. In the linear regime, 
for sufficiently large populations, these fluctuations are gaussian 
and scale as the square root of the population size (Bailey 1975), while 
for smaller population sizes the relative fluctuations are larger and may 
lead to stochastic extinction (Bartlett 1957). More recently, 
several authors have stressed the need for realistic models, both forced and 
unforced, to include stochastic effects (Aparicio \& Solari 2001, N\aa 
sell 
1999, 
Rohani {\it et al.} 1999, 2002) in order to describe various aspects of 
incidence time series, such as the observed different patterns of recurrent 
epidemics.
A more systematic treatment of the shortcomings of deterministic models
was given by Lloyd (2004), who used different analytic techniques as 
well as numerical simulations to investigate the departures from 
deterministic (or mean-field) behaviour of the unforced and seasonally 
forced SIR models, as a function of the population size. 
This analysis highlights serious deficiencies of deterministic descriptions 
of dynamical (long term) behaviour, especially for seasonally forced models 
and 'moderate' size ($< 10^{7}$) populations.

Spatial structure is known to be another essential ingredient of the
dynamics of epidemic spread. Coarse grained metapopulation models have 
been considered to describe phase-locking and decoherence of the observed
spatial patterns of infection, as well as their effects on global 
persistence (Bolker \& Grenfell 1995, Finkenst\v{a}dt \& Grenfell 1998, 
Grenfell {\it et al.} 2001, Lloyd \& May 1996, Rohani {\it et al.} 1999).

The breakdown of the 'perfect mixing' hypothesis at a finer spatial
scale is yet another cause of departure from mean-field behaviour,
namely through the screening of infectives, that has been recognized 
in individually based cellular autamata simulations of the short term
dynamics of epidemic bursts (Keeling 1999, Kleczkowski \& Grenfell 1999, 
Rhodes \& Anderson 1996). Less well known is the fact that spatial 
correlations may greatly enhance the stochastic fluctuations.

In this paper, we perform long term stochastic simulations of individually 
based cellular automata on small world networks with SIR and SEIR node 
dynamics. By contrast to spatially structured metapopulation 
models, based on a coarse grained distribution of the global population 
over a few interacting patches (Bolker \& Grenfell 1995, 
Finkenst\v{a}dt \& Grenfell 1998, Grenfell {\it et al.} 2001, Lloyd \& May 
1996, Rohani {\it et al.} 1999) in our model the nodes of the network 
represent individuals. The links are then connections along which the 
infection spreads. A small world 
network simulates the real network of contacts, where local links predominate 
but social and geographical mobilities imply a fraction of random connections 
through which long-range transmission may occur.

We start by assessing the effect of network topology on the long term dynamics 
of a simple spatially extended model, implementing SIR node dynamics 
on a cellular automaton living on a square lattice with small world 
interaction rules (section 2.B). We find characteristic long term dynamics 
related, in a quantitative fashion, to the structure of the network of 
contacts. In particular, the increase in spatial correlations (i) 
enhances the fluctuations around the endemic state; (ii) decreases the 
effective transmission rate through the screening of infectives and 
susceptibles; and (iii) increases the period of the incidence oscillations 
as a result of the lower effective transmission rate. These changes do not 
follow their mean-field relations revealing the presence of important spatial 
structure (Sections 2.C and 3.B).

In Section 3 we test the model in a more realistic setting, using
SEIR node dynamics, a birth rate of $1/61 year^{-1}$ and the values 
reported in the literature (Anderson \& May 1991) for the latency and recovery 
times of two childhood diseases, measles and pertussis.
For measles, the pre-vaccination pattern is accurately reproduced by 
simple deterministic models with seasonal forcing. This is not the case 
for pertussis, where it is well known that stochasticity plays an important 
role.
A stochastic version of the standard seasonally forced SEIR 
(Susceptible-Exposed-Infective-Recovered) model generates, for pertussis 
in the pre-vaccination era, a temporal pattern consistent with the 
observed dynamics, and for the vaccination era, epidemics with a pronounced 
3.5 year period (Rohani {\it et al.} 1999) in line with the epidemiological 
data. 
These results were obtained by assuming different amplitudes of seasonal 
forcing for measles and for pertussis.

In Section 3.B we consider the incidence time series given by the cellular 
automata when the model independent parameters are taken for measles.
As with any unforced model, the characteristic annual/biennial cycles 
of prevaccination records cannot be obtained without fine tuning of the 
model's parameters, infectiousness and probability of long-range 
infection. However, in the small world region, the amplitudes of
the fluctuations are shown to be compatible with incidence records for 
populations with similar sizes.
In order to compare the infectiousness parameter with the values in the 
literature we have investigated the behaviour of our model in the limit 
where the links between the nodes are completely random.
Apart from providing a way to relate the basic reproductive rate with 
the infectiousness parameter in our model, this analysis reveals the 
mechanism behind recurrent epidemics in unforced spatially extended models,
showing how fluctuactions decrease along with the spatial correlations as 
we approach the homogeneously mixed stochastic model.

In Section 3.C we show how the unforced SEIR model on small world networks 
produces sustained oscillations with periods and amplitudes compatible 
with the pertussis data both before and after mass vaccination for realistic 
values of the three model independent parameters, life expectancy and latency 
and recovery times. Analysis of the homogeneously mixed limit shows that 
the infectiousness parameter is also within the reported range for pertussis 
(Anderson \& May 1991). 

Our results for the SIR and SEIR models support the conclusion 
that, for some purposes, successful modeling of disease spread must take into 
account that populations are finite and discrete, and must include a realistic 
representation of the spatial degrees of freedom or, more generally, of the 
interaction network topology. It also stresses the importance of 
network connections in real epidemics, an ideia that has been acknowledged 
recently in the context of threshold behaviour (May \& Lloyd 
2001, Pastor-Santorras \& Vespigniani 2001). 
Finally, it calls for a reassessment of the impact of environmental forcing 
by studying its effects on a more realistic autonomous model as the one that 
we propose. 
In particular, fluctuation enhancement by spatial correlations might remove 
some of the constraints on the strength of seasonality and avoid fine tuning 
of seasonal forcing amplitudes.  

\section{SIR node dynamics on a small world network}

\subsection{Deterministic and stochastic SIR models}

In order to set the notation we start with a brief description of the 
deterministic and stochastic SIR models.
 
A community of $N$ (fixed) individuals comprises, at time $t$, $S$ 
susceptibles, $I$ infectives in circulation and $R$ recovered or removed 
(isolated, dead or immune). 
Constant infection, $\beta$, recovery and birth/death rates are assumed. 
$\gamma$ is the recovery rate while birth and death rates, $\mu$, are 
assumed to be equal.

The basic differential equations in terms of the densities, $s$ and $i$, 
are:
\begin{equation}
{d s  \over  dt } = - \beta si - \mu s + \mu 
\end{equation}
\begin{equation}
{d i  \over  dt } = \beta si - (\gamma + \mu )i 
\end{equation}

Endemic equilibrium occurs at $i_0= \mu (1/(\gamma + \mu )-1/\beta)$ and 
$s_0= (\gamma + \mu) /\beta$. By linearizing around this point we obtain 
for the period of the damped oscillations, $T \approx {2 \pi \over \sqrt 
{\mu (\beta - \gamma)}}$.

By contrast with some SIR models we have considered non-zero birth 
rates, $\mu$, that will turn out to be crucial to describe the 
long-term dynamical behaviour of the spatial version of the system 
(Sections 2.C and 3).

The stochastic version of the SIR model considers the probability that 
there are $s$ susceptibles and $i$ infectives and three types of 
transitions:
\begin{equation}
Pr\{(s,i)\rightarrow (s-1,i+1)\} = \beta si \Delta t 
\end{equation} 
\begin{equation}
Pr\{(s,i)\rightarrow (s,i-1)\} = \gamma i \Delta t + \mu i \Delta t
\end{equation} 
\begin{equation}
Pr\{(s,i)\rightarrow (s+1,i)\} = \mu \Delta t - \mu s \Delta t
\end{equation} 

The stochastic change in the number of susceptibles in time $\Delta t$ 
is the sum of the loss due to infection, Poisson 
distributed with mean N$\beta si\Delta t$, or death, also Poisson 
distributed with mean N$\mu s \Delta t$, and the gain due to the arrival 
of new susceptibles, Poisson distributed with mean N$\mu \Delta t$. 
Similarly, the stochastic change in the number of infectives is the sum 
of a Poisson distributed gain due to infection, with mean N$\beta si 
\Delta t$, and a loss from death or recovery Poisson 
distributed with mean N$(\gamma + \mu )i \Delta t$.  

The endemic equilibrium is the same as that of the underlying 
deterministic model.
The equations for the stochastic changes in $s$ and $i$ may be linearised 
about the endemic equilibrium values and the variance of the fluctuations 
is easily obtained $\sigma^2_s= 1 /s_0 + (1-s_0)/i_0$ and 
$\sigma^2_i = 1/(i_0(1-s_0)) + (s_0(1-s_0))/i_0^2$. 

We note that in this regime the fluctuations about the endemic state 
scale as the square root of the system size. These fluctuations exhibit 
a temporal structure modulated by the underlying deterministic 
dynamics and thus oscillate (on average) with the period of the damped 
oscillations of the corresponding deterministic model. The amplitude of 
these fluctuations is, however, small and the size
of the recurrent epidemic peaks is often underestimated by a perfectly 
mixed stochastic model.

\subsection{Stochastic SIR model on a small world network}

The SIR model is easily generalized to an epidemic that takes place on a 
network. As for models of epidemic spread on regular lattices (Grassberger 
1983), the model may be mapped onto percolation on the same network 
(Moore \& Newmann 2000a). The percolation transition corresponds to the 
epidemic threshold, above which an epidemic outbreak is possible (i.e. 
one that infects a non-zero fraction of the population, in the limit of  
large populations) and the size of the percolating cluster above this 
transition corresponds to the size of the epidemic.

In recent works, the role of the network topology for SIS and SIR
models with zero birth rate has been considered in the calculation of 
epidemic thresholds (May \& Lloyd 2001, Moore \& Newmann 2000a, 2000b, 
Pastor-Santorras \& Vespigniani 2001a) the stationary 
properties of the endemic state (Pastor-Santorras \& Vespigniani 2001b),
and in the short term dynamics of epidemic bursts (Keeling 1999, 
Kleczkowski \& Grenfell 1999, Rhodes \& Anderson 1996). By contrast, the 
effects of network topology on the long term dynamics of epidemic spread 
remain poorly understood. 

We shall consider a small world contact network. This type of networks have 
topological properties that interpolate between lattices and random 
graphs, and were first proposed by by Watts and Strogatz (1998) as 
realistic models of social networks. 
A fraction of the links is randomized by connecting nodes, with probability 
$p$, with any other node on the lattice; the number of links is preserved by 
removing a lattice link when a random one is established. The interpolation 
is non linear: for a range of $p$ the network exhibits small world behaviour, 
where a predominantly local neighbourhood (as in lattices) coexists with a 
short average path length (as in random graphs). A small world network over 
a regular network is defined, in a statistical sense, by the small world 
parameter $p$. Another useful quantity to characterize it is the 
clustering coefficient $C$, that measures the locality of the network. The 
clustering coefficient is defined as the averaged fraction of all the 
possible links between neighbours of a given node that are actually present 
in the network.
Analysis of real networks (Dorogotsev \& Mendes 2003) reveals the 
existence of small-worlds in many interaction networks, including networks 
of social contacts.

In order to take into account spatial variations ignored in homogeneously 
mixed stochastic models, we consider a cellular automaton (CA) on a square 
lattice of size $N=L^2$. 
The (random) variables at each site may take one of three values: $S$, 
$I$ or $R$. The lattice is full, $N = S+I+R$. We consider local interactions 
with $k$ neighbouring sites, as well as random long-range interactions, with 
a small world probability, $p$, with any other site on the lattice. The total 
rate of infection $\beta$ is the sum of the local and long-range rates of 
infection, given in terms of the small world parameter by $\beta(1-p)$ and 
$\beta p$, respectively. 
We have used different types of boundary conditions and checked that the 
results do not depend on them. 

Birth, death and infection occur stochastically, with fixed rates 
($\mu$, $\mu$, $\beta$) while recovery, characterised by a disease 
dependent period, is deterministic (after $\tau _i ={ 1 /\gamma }$ 
time steps). This disease dependent period sets the time scale of the model. 
At each Monte Carlo step, $N$ random site updates are performed 
following a standard algorithm. One type of event, long and short-range 
infection, death and birth, is chosen with the appropriate rate 
($\beta p$, $\beta (1-p)$, $\mu$, $\mu$, respectively) and then proceeds 
as follows. For infection events, one site is chosen at random; if the 
site is in the infective, $I$, or recovered, $R$, states no action is taken. 
If the site is in the susceptible state, $S$, one other site (one of its 
$k=12$ nearest neighbours for local infection or any other site on the lattice 
for long-range infection) is chosen at random; the first site becomes infected 
iff the second site is in the infective state, $I$. 
For a death event, a site is chosen at random and death occurs regardless; 
the state is then changed into the removed or recovered state, $R$. 
Finally, for birth events a recovered site, $R$, is searched at random until 
it is found; then it is changed into the susceptible, $S$ state.
 
To account for the deterministic recovery, counters are assigned to all 
sites and are set to -1; the counters of the infective sites, 
$I$, are updated at each time step. After $\tau _i = {1 / \gamma }$ MC 
steps (recovery time) the infectives recover, that is the sites in the $I$ 
state change into the $R$ state.

\subsection{Results for the Stochastic SIR model on a small world network}

Apart from the network parameter $p$ that determines the fraction
of long distance infection, the spatial SIR model depends only on three 
additional parameters. Two are the infectious period and life expectancy, 
and the other, $\beta$, the infectiousness of the disease. The 
first two are determined from medical and demographic data (Anderson \& 
May 1991) and we show that $p$ and $\beta$ may be obtained from fits to 
epidemiological data. Thus the model provides quantitative information on the 
structure of the network of contacts underlying the spread of a particular 
disease.

In Figure 1 we summarize the results for the persistence (fraction of the
simulations surviving a given number of time steps) as a function of the
small world parameter $p$ for a population of $N= 250 000$.
The other model parameters are kept fixed at $\mu = 0.0006$ day$^{-1}$, 
$\gamma = 0.0625 = {1 / 16}$ day$^{-1}$ and $\beta = 0.66$ day$^{-1}$. 
The persistence is almost zero at low $p$, and approaches one over a narrow 
range of $p$, for a population of this size.
The transition between extiction and persistence is a percolation 
transition. The epidemic persists in finite populations as a 
result of the finite birth rate that allows the renewal of susceptibles. 
 
On regular lattices the transitions are isotropic percolation 
transitions if a site cannot be re-infected and directed percolation if 
re-infection (through birth of susceptibles, mutation of infectious 
agents, etc.) occurs (Dammer \& Hinrichsen 2003). In the 
presence of random long-range interactions on the complex network 
the transitions become mean-field with an anomalous asymptotic region 
(Hastings 2003). In addition, finite size effects broaden the transition 
region considerably. We note that by contrast with atomic systems, whose 
sizes are effectively infinite, finite size effects cannot be ignored in this 
context. A detailed study of these effects, as well as the dependence of the 
thresholds on relevant epidemiological parameters, will be published elsewhere. 

In Figure 2 we have plotted the 
clustering coefficient (Watts \& Strogatz 1998) of the underlying small 
world network for the parameters of Figure 1. 
The persistence and the root mean square amplitude of the epidemic peaks 
of the SIR model are also shown. 
When the clustering is large (as in lattices) the fluctuations are 
large and stochastic extinction occurs. When p=1 the fluctuations are 
small (as in perfectly mixed stochastic models) and the epidemic 
persists. The persistence transition occurs in the region of intermediate 
clustering, i.e., at the edge of the small world regime where the 
fluctuations are sufficiently small for stochastic extinction to 
become rare.

The effective transmission rate, $\beta _{eff}N$, the average number of 
new infectives per time step divided by the product of the instantaneous 
densities of susceptibles and infectives, is shown in Figure 3. 
These rates differ from those reported by Kleczkowski and Grenfell (1999) 
for simulations spanning a single epidemic wave, in two ways. 
First, we focus on the long term dynamical regime and exclude from the 
analysis the transient corresponding to the initial dynamical regime. In 
addition, the effective transmission rates are calculated during the 
simulation runs while those reported by Kleczkowski and Grenfell (1999) 
were obtained by fitting simulated time series to approximate mean-field 
solutions.

The significant variation of the effective transmission rate with 
$p$ is due to the clustering of infectives and susceptibles and has to be 
taken into account in fittings to effective mean-field models.
This clustering, and spatial correlations in general, have also
drastic consequences on the amplitude and on the period of the incidence 
oscillations. The typical values of these quantities for a
given time series can be used for model fitting and testing. We will 
return to this point in the next section.

\section{SEIR node dynamics on a small world network}

\subsection{Stochastic SEIR model on a small world network}

In this section, we show that epidemic models on small world
networks provide a natural basis for the description of recurrent
epidemic dynamics, by reporting realistic modelling of two childhood 
diseases, measles and pertussis. 
We have obtained, using the SEIR model, sustained oscillations with 
periods and amplitudes compatible with the data for realistic values of 
the model parameters. 
In recent works (Johansen 1996, Kuperman \& Abramson 2001) a relation was 
suggested between spatial structure and the onset of recurrent epidemics, 
but in both models the periods of the incidence oscillations are of the 
order of the time elapsed between infection and loss of immunity, and thus 
the conclusions are not relevant for immune for life diseases. 

The spatial SEIR model is similar to the spatial SIR model, but 
the recovery time is split into latent and infectious periods. During the 
latent period an $S$ has become infected but cannot yet infect another 
susceptible.
The rules are as in the spatial SIR model with the following modifications.
Infection: $S$ is infected iff the second site is an infective in the 
infectious period, $I$; in this case $S$ is changed into the exposed $E$ 
state.
The counters of the exposed and infective sites are updated. After 
$\tau_l$ steps (latency) the exposed sites, $E$, change into the infective 
state, $I$; the infectious period lasts for an additional $\tau_i$ steps. 
Recovery is deterministic and occurs after $\tau_l +\tau_i$ steps.

We have taken $\mu = 1/61$ year$^{-1}$ for the birth rate and a population 
size of $N= 10^6$ in all the simulations. For measles, we have set
$\tau_i = 8$ days, $\tau_l = 6$ days and $\beta = 2.4$ day$^{-1}$ for 
the homogeneously mixed model, $p=1$, and $\beta = 4.75$ day$^{-1}$ 
for the network model with $p=0.2$. 
For pertussis we set $\tau_i = 18 $ days, $\tau_l = 8$ days and 
$\beta = 1.5$ day$^{-1}$ for the homogeneously mixed model, $p=1$, and 
$\beta = 4.0$ day$^{-1}$ for the network model with $p=0.2$. Given the 
length and variability of the infectious period of pertussis, 
deterministic recovery was changed for Poisson recovery. The end of the 
latency period was kept deterministic.

\subsection{Results for SEIR dynamics simulations of measles}

We obtained new infectives time series on $N=1000\times 1000$ lattices. 
The time series for measles in homogeneously mixed populations, $p=1$, 
exhibit sustained fluctuations (Figure 4a) with an average period of two years. 
However, the amplitude of the fluctuations is underestimated when 
compared with the incidence oscillations of measles in 
Birmingham (From data at 
http://www.zoo.cam.ac.uk/zoostaff/grenfell/measles.htm) (Figure 4c) 
with a similar population size.

As discussed in the previous section we found that the fluctuations are 
enhanced as a result of clustering when $p$ decreases. Consequently, in a 
range of $p$ around the persistence transition we can account for the 
amplitude of the incidence peaks, in sharp contrast with homogeneously mixed 
stochastic models. Conversely, reliable values of the incidence peaks become 
in view of this a measure of $p$, leaving only $\beta$ to be fitted to 
experimental data.

In Figure 4b we plot new infectives time series, for SEIR simulations on a
network with $p=0.2$, and $\beta$ corresponding to incidence oscillations 
with an average period of two years (other parameters as in
Figure 4a). The amplitude of the incidence peaks is shown to increase
significantly, in line with the real data. 
As a further check we report the average age at infection for both models. 
We found that while the homogeneously mixed model underestimates the 
average age at infection ($\sim 3$ years) the small world model, with 
$p=0.2$, increases it by $20$\% in line with epidemiological data. Since 
the incidence oscillations have similar periods in both models this result 
provides additional support of the network model. 

We remark that the effect of spatial correlations on the period is
striking. We found that the period increases by up to a factor of two as $p$
decreases, as a consequence of the reduced effective transmission rate, 
sugesting that the contact network structure may play a role in the 
triennial epidemic cycles observed in the pre-vaccination measles records of 
cities such as Copenhagen and Baltimore (Bolker \& Grenfell 1995). 
However, spatial correlations have further implications, beyond the screening 
of infectives and its consequences. We have shown that stochastic fluctuations 
are significantly enhanced, and that, as the homogeneous mixing relations 
break down, the period and the average age at infection change 
independently in this model.

Despite the fact that, for measles, seasonality cannot be ignored, and that 
pre-vaccination incidence records are well reproduced by simpler deterministic 
models with seasonal forcing, the reported effects of spatial correlations 
should also be taken into account. We expect that adding seasonality 
to our model will produce time series showing robust annual/bienneal peaks 
with the observed incidences for a wide range of forcing amplitudes. 

\subsection{Results for SEIR dynamics simulations of pertussis}

While for measles seasonal forcing is the basic ingredient to explain
pre-vaccination long term dynamics, this is not the case for pertussis,
where stochasticity plays an important role.

We have obtained, using SEIR simulations, new infectives time series on 
$N=1000\times 1000$ lattices for pertussis. Given that the infectious 
period of pertussis varies from 14 to 21 days we modelled the recovery of 
pertussis stochastically. 
In Figure 5a we plot new infectives time series for pertussis on the same  
network as measles ($p=0.2$), with $\beta=4.0$ day $^{-1}$ corresponding 
to pre-vaccination incidence oscillations with a broad 2.5 year period. The 
root mean square amplitude of the incidence peaks is compatible with reported 
epidemiological data (Rhoani {\it et al.} 1999). 
It is obvious from Figures 4 and 5 that pertussis exhibits epidemic peaks 
that are much noisier than those of measles. In our model this is due in 
part to the stochastic recovery process. However, the effects of spatial 
correlations, that are significant close to the epidemic persistence 
transition, are much less evident for pertussis. 
This is confirmed by the results for pertussis of the homogeneously 
mixed SEIR model, $p=1$, that are plotted in Figure 5b. This time series 
exhibits sustained fluctuations (Figure 5b) with an amplitude that 
is similar to that of the network model ($p=0.2$), in marked contrast 
with the results for measles (Figure 4) for the same contact network. 
Therefore, the importance of the spatial correlations depends also on the 
particular disease.

Finally, in Figure 5 we have plotted new infective time series for 
pertussis after mass vaccination. We assumed a vaccination coverage 
of $80\%$ with $70\%$ efficacy and modelled it by reducing the 
birth rate to one-half of its value in the pre-vaccination era. It is 
clear that the unforced spatial SEIR model is capable of describing the 
increase in the period of the recurrent 
epidemic peaks reported after mass vaccination while increasing their 
spatial coherence, in agreement with available data
(Rohani {\it et al.} 1999). In our model, this increased coeherence is 
due to the loss of stability of the endemic equilibrium.

These results show that spatially extended SEIR models on small world 
networks exhibit sustained oscillations, that differ from the stochastic 
fluctuations of perfectly mixed stochastic models, under various 
conditions relevant to the description of childhood diseases. In 
particular, both the observed periods and the spatial coherence of the 
epidemic peaks of pertussis, before and after mass vaccination, are 
described by this spatially extended unforced model.
 
\section{Discussion and Conclusions}

We performed long term stochastic simulations of individually based
cellular automata on small world networks with SIR and SEIR node
dynamics.
This model has the recognized basic ingredients for realistic modeling, 
namely stochasticity, spatial structure given by a plausible
contact network, a discrete and reasonably sized finite population, and
other parameters as given in the literature for two childhood diseases.

As expected, the long term dynamics is dominated by stochastic fluctuations
around mean-field behaviour. However, when the network's clustering 
coefficient is in the small world region, i.e., in the range of typical 
values measured for social networks, the amplitude of the stochastic 
fluctuations is enhanced significantly by spatial correlations. In 
particular, away from the boundary of the small world region, where the 
clustering coefficient is high and the interactions are predominantly 
local, fluctuations lead to extinction on short time scales,
for other parameters above the endemic threshold of the homogeneously 
mixed model. For the remaining range of the clustering coefficient,
epidemics persist for times that approach those of perfectly mixed 
populations providing suitable models to investigate the effects of 
spatial correlations on the long term dynamics of epidemic spread.

This has three major consequences. First, when spatial strucuture is taken 
into account, large populations will have to become even larger to average 
out significant differences from the deterministic model's behaviour. Second, 
for small world networks, the persistence threshold corresponds to the region 
where the clustering coefficient starts to decrease, along with relative
fluctuations.
Third, time series of the (unforced) model provide patterns of recurrent 
epidemics with slightly irregular periods and realistic amplitudes, 
suggesting that stochastic models together with spatial correlations are 
sufficient to describe the long term dynamics when seasonal forcing is 
weak or absent. 

Annual and biennial regular temporal patterns are the signature of 
seasonally forced models, but these may of course exhibit different 
dynamics, such as triennial cycles and chaotic behaviour. 
Stochastic effects in seasonally forced models can also produce 
slightly irregular, realistic, time series. 
Is there then a way to determine the basic mechanism of recurrent 
epidemics from the case report data ?
Our model reduces to the mean-field stochastic model for p=1 and large N, 
but in the range of epidemiologically relevant population sizes, finite 
size effects are important and spatial correlations cannot be ignored.
The epidemic peaks obtained from our model scale as $N^a$, $1/2 \leq a < 
1$, $a-1/2$ being a measure of the effect of spatial correlations.  
By contrast, seasonally forced peaks scale as $N$. 
For a given disease and a given seasonality, the peaks over several 
cities must scale with the population $N$.

The different scaling forms of epidemic peaks predicted by the two 
models could be used to discriminate the underlying mechanism of the 
incidence oscillations characteristic of a given class of diseases.
Linear fits for measles in England and Wales do rather well but 
two-parameter power law fits do almost as well with $a \approx 3/4$. 
However the importance of seasonal forcing for this example is unquestionable,
and a sublinear fit may be the result of a de-phasing effect.  

In conclusion, we have shown that fine grained discrete models based on
contact structure are realistic autonomous models, that are computationally
feasible for relevant population sizes. In the small world regime, they
exhibit sustained fluctuations with well defined period and amplitudes
within the observed range. These quantities may be used to fit the model's 
free parameters, $p$ and $\beta$. The fact that this can be done with 
satisfactory quantitative agreement for the simplest version of the model 
puts it on the level of the traditional deterministic and stochastic 
models as a basis for realistic modelling. The impact of other 
factors, such as seasonal forcing, age structure and spatial heterogeneity, 
should be reassessed based on this.

\section*{Acknowledgements}

Financial support from the Portuguese Foundation for Science and 
Technology (FCT) under contract POCTI/ESP/44511/2002 is gratefully 
acknowledged. JAV acknowledges support from FCT through grant no
XXXXXXXXX.

\newpage

\centerline {FIGURES}

{\bf FIGURE 1} Epidemic persistence transition. Fraction of the 
simulations that survive for 20,000 days as a function of the small world 
parameter $p$, for a discrete SIR model. Infection occurs between 
connected sites. Each site has $12$ connections, an average fraction $1-p$ 
of which are neighbours on the lattice; the remaining fraction $p$ is chosen 
randomly at each time step. The persistence is almost zero, at low $p$ 
($p<0.07$) and approaches one over a narrow range of $p$ close to $0.09$, 
for a population of 250,000. The abrupt change in persistence is 
a percolation transition on the small world network. 

\bigskip
{\bf FIGURE 2} Small world parameters and the epidemic persistence 
transition. 
Clustering coefficient (dashed line) for the underlying small world network.
The fraction of the simulations that survive for 20,000 days (circles), 
and the root mean square amplitude of the epidemic peaks (dotted 
line) is also shown. The clustering coefficient and the root mean square 
amplitude of the peaks were measured relative to the lattice values.
The persistence transition occurs at the edge of the small world regime.

\bigskip

{\bf FIGURE 3}
Effective transmission rate vs the network parameter, $p$. The effective 
transmission rate, $\beta_{eff}N$, is calculated as the average 
of number of new infectives per time step divided by the product of the 
instantaneous densities of susceptibles and infectives. 
The drastic reduction in $\beta_{eff}N$ is due to the clustering of 
infectives and susceptibles 
as $p$ decreases and the spatial correlations increase. 
The model and the parameters are the same as in Figure 1. 

\bigskip

{\bf FIGURE 4}
New infectives time series for homogeneously mixed and spatially 
structured populations. Number of new infectives every 2 weeks, from SEIR 
simulations on $N = 1000 \times 1000$ lattices. 
(a) The results for measles in a homogeneously mixed population, $p=1$, 
and 
transmission rate $\beta= 2.4$ day $^{-1}$, exhibit sustained flutuactions 
with an average period of two years. The amplitude of the fluctuations 
is underestimated when compared with the amplitude of measles 
incidence peaks from Birmingham (c).
(b) The results for measles on a network with $p=0.2$ and $\beta=4.75$ 
day$^{-1}$ 
exhibit an average period close to two years but the amplitude of the 
incidence oscillations is larger in line with the real data (c).  

{\bf FIGURE 5}
New infectives time series for homogeneously mixed and spatially 
structured populations. Number of new infectives every 2 weeks, from SEIR 
simulations on $N = 1000 \times 1000$ lattices. 
(a) The results for pertussis on a network with $p=0.2$ and $\beta=4.0$ 
day$^{-1}$. The time series is noisier than that for measles on the same 
network. After mass vaccination both the period of the incidence oscillations 
and the spatial coherence increase.  
(b) The results for pertussis on a homogeneously mixed population, $p=1$, 
and transmission rate $\beta= 1.5$ day $^{-1}$, exhibit sustained 
fluctuations with the same average period. The behaviour of pertussis in 
the pre and post-vaccination periods is similar to that of the network model
indicating that stochastic effects dominate the behaviour of these time 
series.

\clearpage
\includegraphics{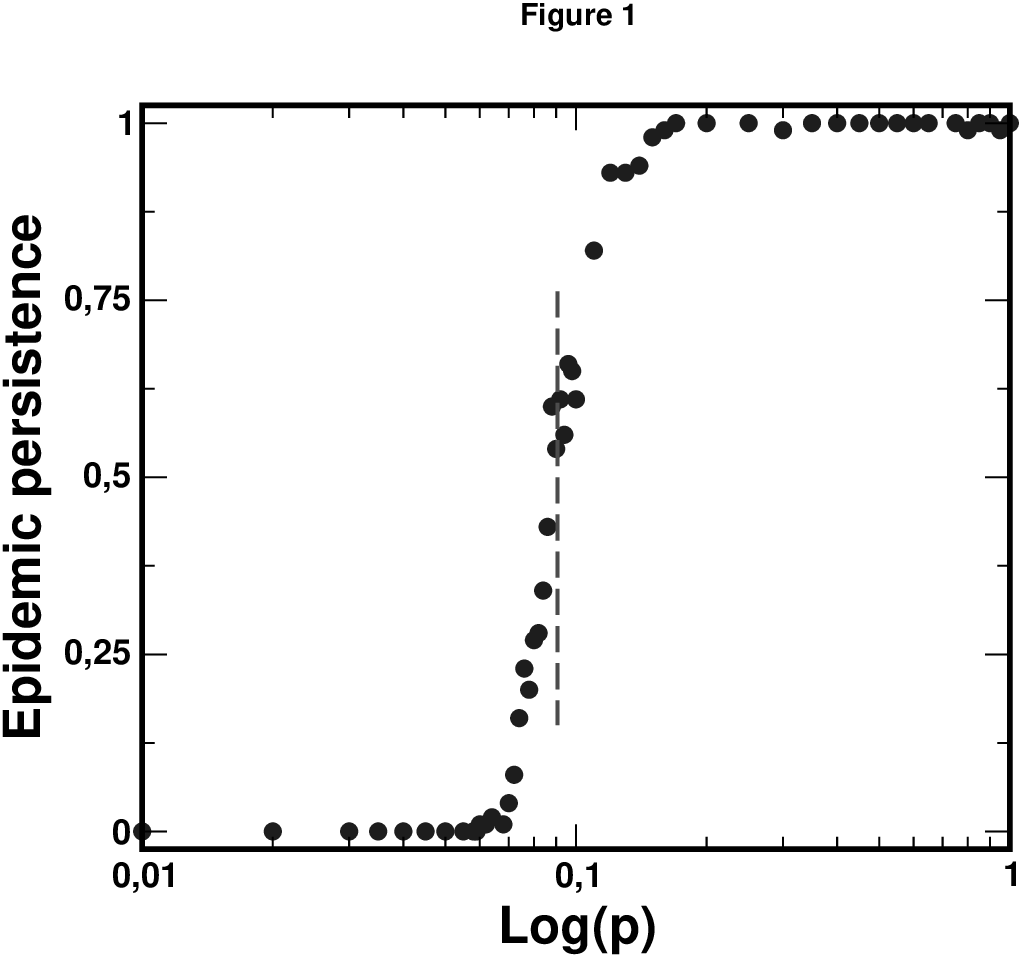}
\clearpage
\includegraphics{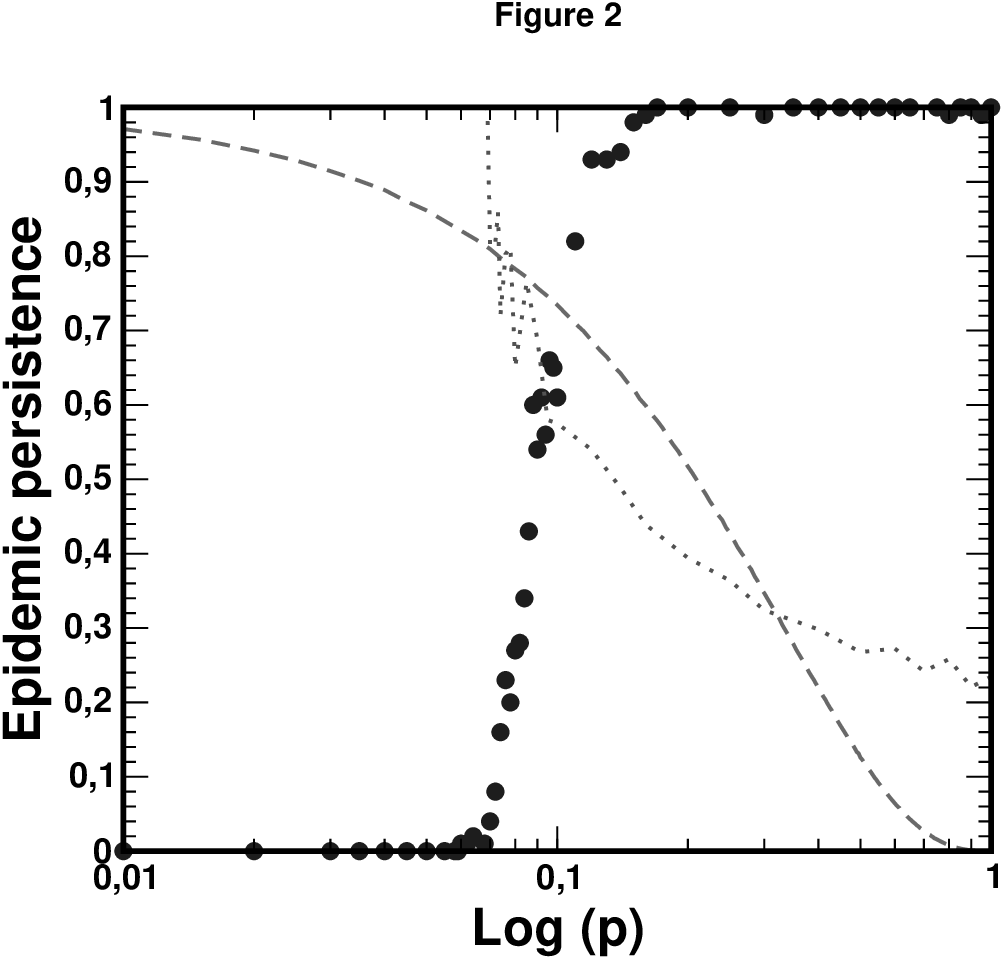}
\clearpage
\includegraphics{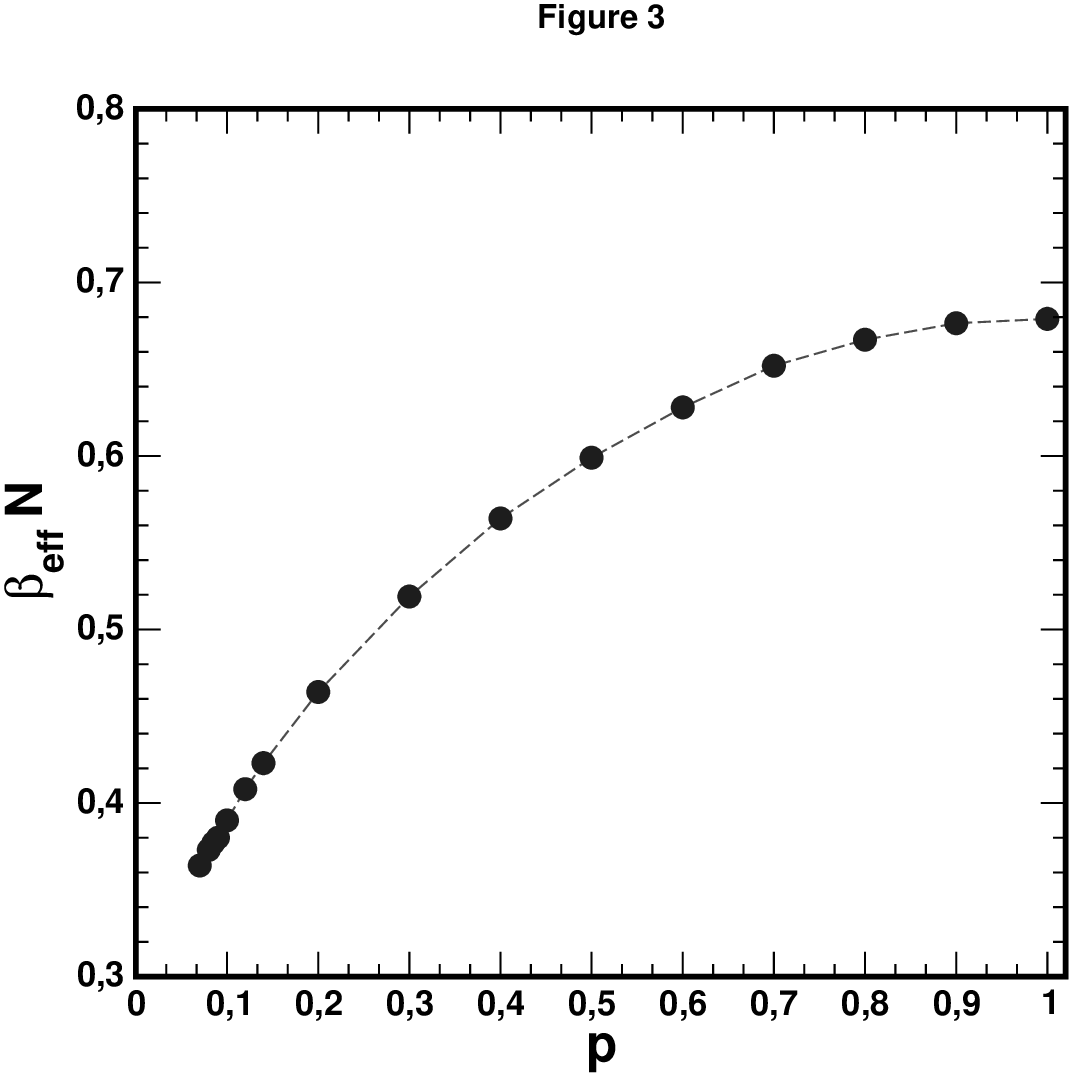}
\clearpage
\includegraphics{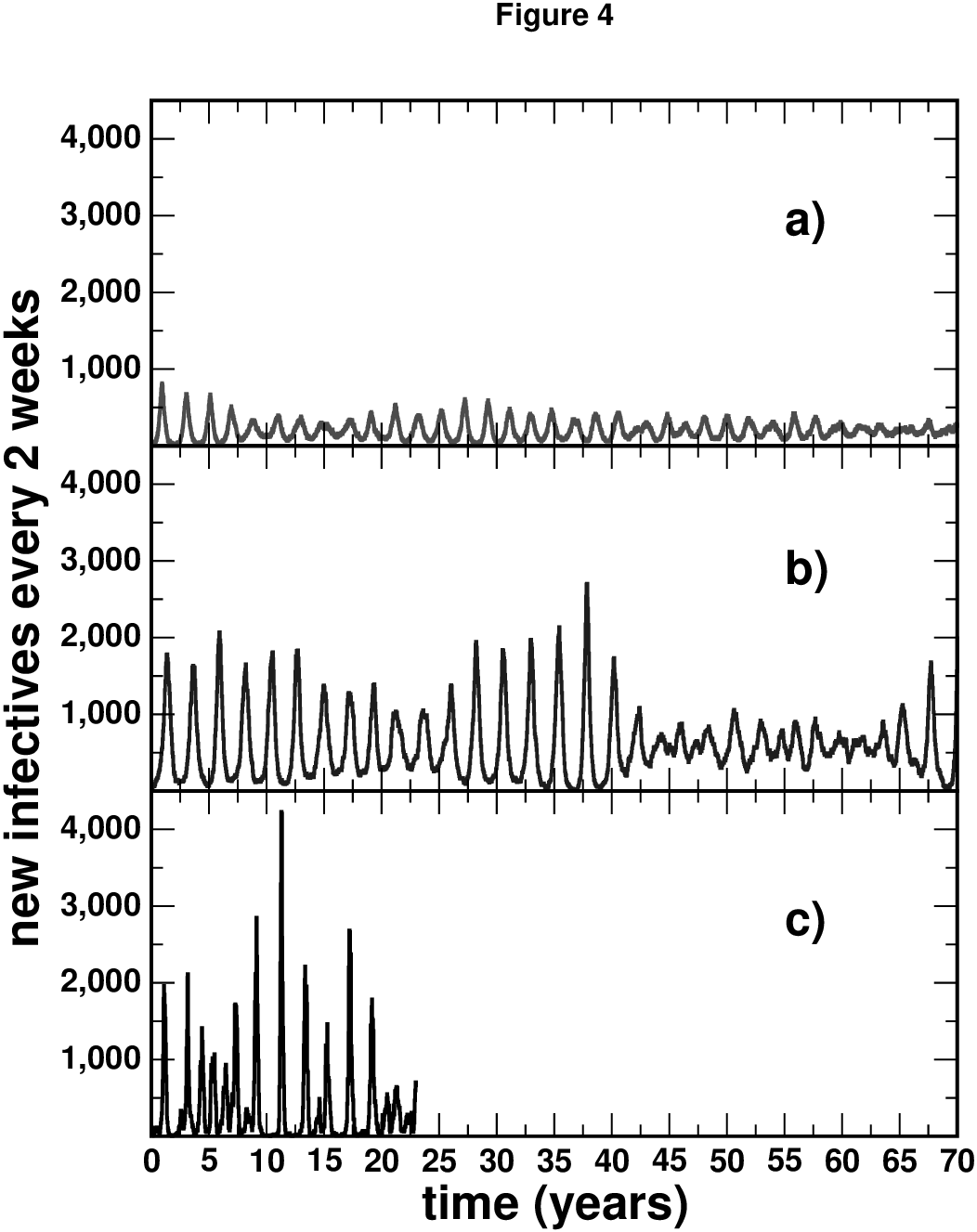}
\clearpage
\includegraphics{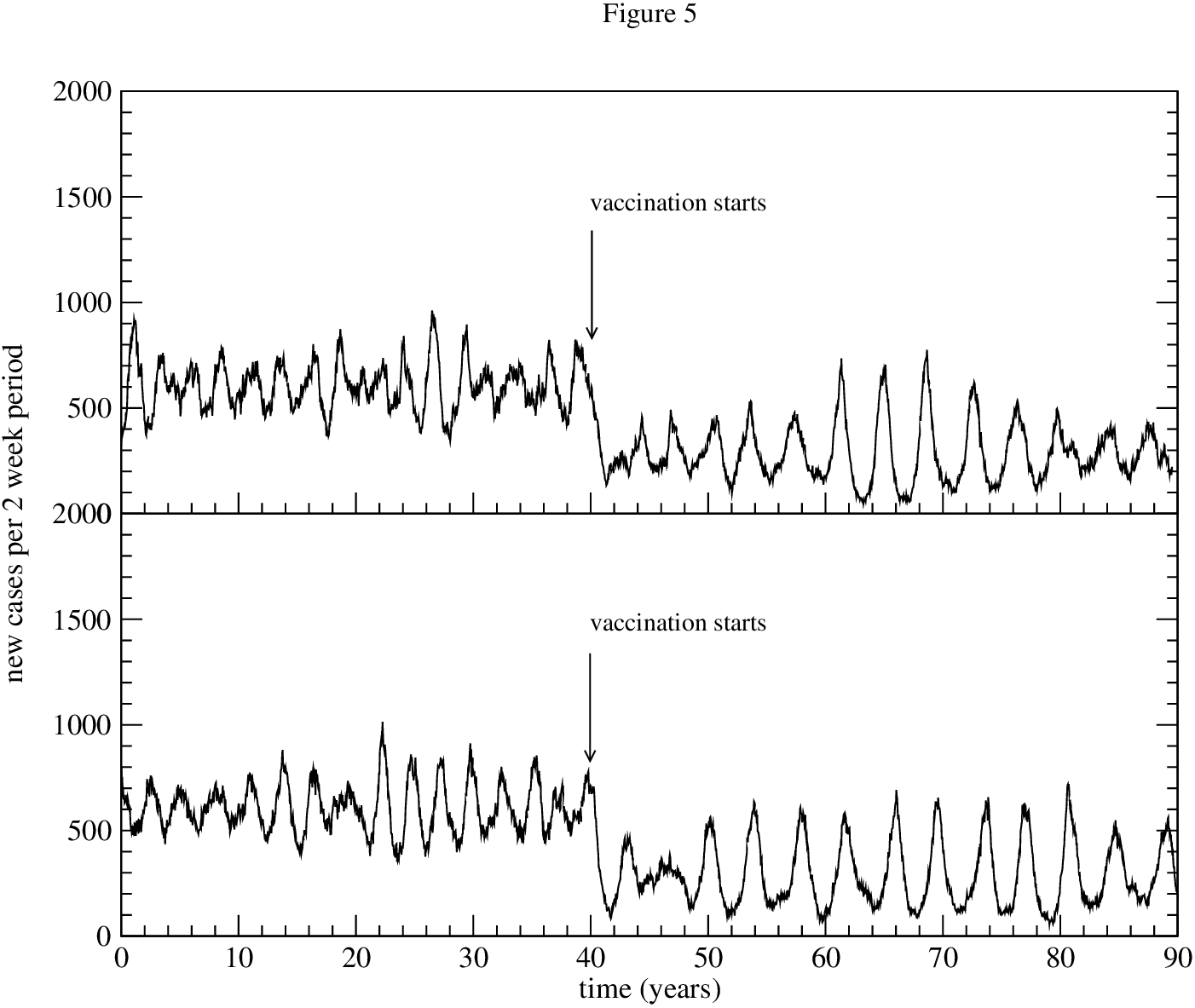}
\end{document}